\documentclass[11pt]{article}
\usepackage[left=1.5cm,right=1.5cm]{geometry}
\usepackage[numbers,square]{natbib}
\usepackage[T1]{fontenc}
\usepackage{tikz-cd}
\usepackage[british]{babel}
\usepackage{layout}
\usepackage{amsmath}
\usepackage{amssymb,amstext, amsthm, simplewick, amsfonts}
\usepackage{framed}
\usepackage{amsfonts}
\usepackage{color} 
\usepackage{braket}
\usepackage{multicol} 
\usepackage{epsfig}
\usepackage{cancel}
\usepackage{caption}
\usepackage{hyperref}
\usepackage{epsf}
\usepackage{feynmf} 
\usepackage{frontespizio}
\usepackage{rotating}
\usepackage{subfigure}
\usepackage{pstricks}
\usepackage{type1ec}
\usepackage{lettrine}
\usepackage{bbold}
\usepackage{calligra}
\usepackage{tikz}
\usepackage{float}
\usepackage{subfigure}
\usepackage{slashed}
\usepackage{mathrsfs}
\usepackage{curve2e}
\usepackage{setspace}
\usepackage{indentfirst}
\usepackage{emptypage}
\usepackage{relsize}
\usepackage{mathrsfs}
\usepackage{stackengine}
\usepackage{calc}
\usepackage{hyperref}
\hypersetup{
	colorlinks,
	citecolor=green,
	filecolor=black,
	linkcolor=blue,
	urlcolor=black
}
\newcommand{\3}[1]{C_{
		\ifthenelse{\equal{\ThreePt}{\empty}}{#1}{
			\ifthenelse{\equal{#1}{\empty}}{\ThreePt}{\ThreePt,#1}}}}
\newcommand{\redef}[1]{{C'}_{
		\ifthenelse{\equal{\ThreePt}{\empty}}{#1}{
			\ifthenelse{\equal{#1}{\empty}}{\ThreePt}{\ThreePt,#1}}}}
\newcommand{\ren}[1]{C_{
		\ifthenelse{\equal{\ThreePt}{\empty}}{#1}{
			\ifthenelse{\equal{#1}{\empty}}{\ThreePt}{\ThreePt,#1}}}}
\newcommand{\sd}[1]{D_{
		\ifthenelse{\equal{\ThreePt}{\empty}}{#1}{
			\ifthenelse{\equal{#1}{\empty}}{\ThreePt}{\ThreePt,#1}}}}

\numberwithin{equation}{section} 
\makeatletter
\@addtoreset{equation}{section}
\newcommand{\bea}{\begin{eqnarray}}
\newcommand{\eea}{\end{eqnarray}}
\makeatother
\newcommand{\beqa}{\begin{eqnarray}}
	\newcommand{\eeqa}{\end{eqnarray}}

\newcommand{\bann}{\begin{eqnarray*}}
\newcommand{\eann}{\end{eqnarray*}}

\newcommand{\bmi}{\begin{minipage}}
\newcommand{\emi}{\end{minipage}}

\newcommand{\be}{\begin{equation}}
	\newcommand{\ee}{\end{equation}}

\newcommand{\beq}{\begin{equation}}
	\newcommand{\eeq}{\end{equation}}
\newcommand{\figref}[1]{Fig.~\ref{#1}}			

\newcommand{\ThreePt}{\empty}
\usepackage{accents}
\makeatletter
\newcommand{\xLine}[2][]{\ext@arrow 0359\Rightarrowfill@{#1}{#2}}
\makeatother
\xdefinecolor{darkgreen}{RGB}{102, 204, 70}
\xdefinecolor{darkblue}{RGB}{0, 0, 153}
\usepackage{amssymb}
\usepackage{pifont}

\usepackage{tikz-feynman}
\usetikzlibrary{arrows}
\usetikzlibrary{snakes}
\usetikzlibrary{decorations,decorations.pathmorphing,decorations.markings}

\tikzset{graviton/.style={decorate, decoration={snake}, double}}
\tikzset{gluon/.style={decorate, decoration={coil, segment length=8, aspect=1.2, amplitude=3 }}}

\begin{document}
	\begin{center}
		\vspace{1.5cm}
	\begin{center}
	\vspace{1.5cm}
	{\Large \bf Semiclassical Lensing and Radiative Lens Equations}\footnote{Presented by C. Corian\`o at the XVII Marcel Grossmann Meeting, Pescara, Italy, 7-12 July 2024} 
	 
\vspace{0.3cm}
	
		\vspace{1cm}
		{\bf $^{(1)}$Claudio Corian\`o, $^{(1),(2)}$ Mario Cret\`i and $^{(1)}$Leonardo Torcellini\\}
		\vspace{1cm}
{\it  $^{(1)}$Dipartimento di Matematica e Fisica, Universit\`{a} del Salento \\
and INFN Sezione di Lecce,Via Arnesano 73100 Lecce, Italy\\
National Center for HPC, Big Data and Quantum Computing\\}
{\it $^{(2)}$Center for Biomolecular Nanotechnologies,\\ Istituto Italiano di Tecnologia, Via Barsanti 14,
73010 Arnesano, Lecce, Italy\\}
\vspace{0.5cm}

	\end{center}
	\begin{abstract}

We analyze lensing of photons and neutrinos in a gravitational field, proposing a method to include radiative effects in classical lens equations. The study uses Schwarzschild and a Reissner-Nordstrom metrics expanded at second post Newtonian order in the Newtonian potential, employing a semiclassical approach to compare one-loop corrections from the Standard Model with Einstein's deflection formula via an impact parameter representation. We also explore the energy dependence of deflection due to quantum corrections and integrate these with classical lens equations.

\end{abstract}
	\end{center}
	\newpage

\section{Introduction}\label{intro}
Classical photon deflections  can be compared at classical and quantum levels by matching the classical gravitational cross section, expressed via the photon's impact parameter, with the perturbative cross section that includes radiative corrections \cite{Coriano:2014gia, coriano2015neutrino,Coriano:2013iba}. This comparison, leads to a differential equation for the beam's impact parameter, linking the classical and quantum descriptions, as proposed in \cite{Delbourgo:1973xe,Berends:1974gk}. Specifically, the energy dependence introduced at one-loop order allows for a new formula that connects the impact parameter \( b_h\equiv b/(2 G M) \), measured in units of the Schwarzschild radius ($2 G M$, the horizon scale) to the beam's energy \( E \) and the angle of deflection \( \alpha \). This energy dependence, absent in Einstein's classical formula, influences all lensing observables, including magnifications, cosmic shears, microlensing light curves, and Shapiro time delays. This indicates that radiative corrections lead to a violation of the classical equivalence principle in General Relativity, a principle that is inherently classical and conflicts with quantum mechanics due to the Heisenberg uncertainty principle.\\
In this framework, gravity is treated as an external background, and the quantum transition amplitude involves the \( TVV \) vertex (where \( T \) is the Standard Model's energy-momentum tensor and \( V \) the electromagnetic current) for photons \cite{Coriano:2014gia}, or the \( Tff \) vertex \cite{Coriano:2013iba,coriano2015neutrino} (with \( f \) representing a neutrino) for fermions. Numerical comparisons between the classical and semiclassical deflection formulas reveal that the energy dependence of the bending angle, while small, becomes more significant at higher energies due to the logarithmic growth of electroweak corrections \cite{Coriano:2014gia, coriano2015neutrino}.\\
The aim of this approach is to propose a method for integrating quantum effects into conventional lens equations and to illustrate the applicability of this method through a comprehensive numerical study. Photon scattering and gravitational lensing have been pivotal areas of research in astrophysics and cosmology, especially for testing Einstein's General Relativity (GR) and investigating the distribution of matter, including dark matter, in the universe 
\cite{Schneider1992}. In extreme scenarios, such as when photons travel near a black hole, the deflections can become highly significant, sometimes even causing photons to orbit the black hole before escaping. This leads to intricate lensing effects and the formation of multiple images. These phenomena have been rigorously studied across various spacetime geometries, including those of Schwarzschild, Reissner-Nordström, and rotating black holes at classical level.\\
While classical photon deflection by gravity is well understood, the impact of full electroweak corrections had not been thoroughly explored before \cite{Coriano:2014gia, coriano2015neutrino,Coriano:2013iba}. As just mentioned, previous studies have focused on QED corrections in weak lensing, but the electroweak corrections, though small, become significant at high energies or near massive gravitational sources. This is particularly relevant for very high energy gamma rays.

\section{The method} 
The approach of \cite{Delbourgo:1973xe,Berends:1974gk} leads to a differential equation that links the cross section defined in terms of the impact parameter $b$, to the deflection angle $\alpha$, aligning well with Einstein's predictions for weak lensing. The fundamental relation  
\begin{equation}
\frac{b}{\sin\theta} \, \vline  \frac{d b}{d\theta}\vline = \frac{d \sigma}{d\Omega},
\label{semic}
\end{equation}
where  $d \sigma/d\Omega$ is computed perturbatively at quantum level and $\theta$ is the scattering angle of the quantum cross section, can be viewed as a differential equation satisfied by the classical impact parameter. 
The solution of (\ref{semic}) takes the general form 
\begin{equation}
b_h^2({\alpha})=b_h^2(\bar{\theta}) +2\int_{\alpha}^{\bar{\theta}} d\theta' \sin\theta' \frac{d \tilde\sigma}{d\Omega'}, 
\label{intg}
\end{equation}
with $b_h^2(\bar{\theta})$ denoting the constant of integration. 
The semiclassical scattering angle $\alpha$ is obtained from (\ref{intg}) as a boundary value of the integral in $\theta$ of the quantum cross section. As discussed in \cite{Coriano:2014gia}, the integration constant derived from (\ref{intg}) has to be set to zero for $\bar\theta=\pi$, in order for the solution of (\ref{semic}) to match the classical GR $\alpha\sim 2/{b_h}$ at very large $b_h$.\\
When radiative corrections are included, the deflection angle becomes energy-dependent, creating a "gravitational rainbow." This phenomenon, absent in both classical and quantum calculations at the Born level, highlights a key difference between the two approaches. Though its impact is typically small, it becomes significant near the event horizon of a massive and very compact object.\\
The formalism employs a retarded graviton propagator, ignoring the back-reaction of the scattered beam on the source, similar to a typical scattering problem with a static external potential. Due to the presence of a horizon, we identify a lower bound on the impact parameter where classical General Relativity (GR) and quantum predictions align. This bound is around 20 \( b_h \), close to the horizon of the source, where the two descriptions agree.
For smaller impact parameters ($4 < b_h  < 20$), discrepancies arise between the approaches (see the analysis in \cite{Coriano:2014gia, coriano2015neutrino}). As the beam nears the photon sphere, the logarithmic singularity in the deflection angle becomes significant, reflecting the limitations of the weak field approximation used in the metric.

\subsection{The neutrino case}
In the neutrino case the one-loop cross section in the electroweak theory is computed from the $Tff$ vertex, with one graviton and two external fermion lines, giving

\begin{figure}[t]
\centering 
\includegraphics[width=0.4\textwidth]{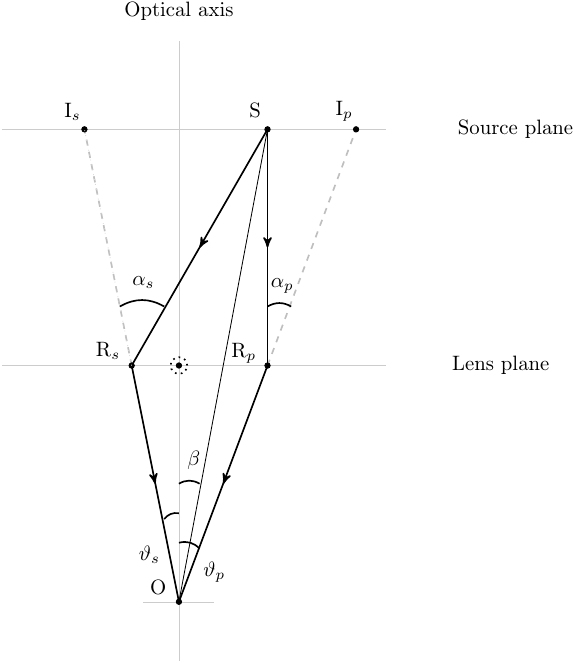} 
\caption{Lens geometry, showing the primary I$_p$ and secondary I$_s$ images generated by the two geodesics of the isotropic emission from the source plane. $D_{LS}$ denotes the distances between the lens and the source planes, and $D_{OS}$ the distance between the observer and the source, measured along the optical axis of the lens.  }
\label{geo1}
\end{figure}

\begin{equation}
\label{sigmaOL}
\frac{d \sigma}{d \Omega}=G^2M^2\frac{\cos^2\theta/2}{\sin^4\theta/2}\left\{1+\frac{4\,G_F}{16\,\pi^2 \sqrt{2}}  \left[ \, f_W^1(E,\theta) + f_Z^1(E,\theta) - \frac{1}{4} \Sigma_Z^L - \frac{1}{4} \Sigma_W^L\right] \right\},
\end{equation}
whose explicit expression has been given in \cite{coriano2015neutrino}. $G_F$ is the Fermi constant, while 
$f_W\, f_Z,  \Sigma_W, \, \Sigma_Z$ are computed perturbatively and denote contributions generated by the exchange of  $W^\pm$ and $Z$ gauge bosons in the loops. 
For massless neutrinos, loop corrections do not induce flavor transition vertices, such as those computed in  \cite{Coriano:2013iba}.
In the case of a point-like gravitational source and of neutrino deflection, at Born level one obtains from (\ref{intg}) and \eqref{sigmaOL} the differential equation
\begin{eqnarray}
\frac{d b^2}{d\theta}&=&- 2  \left(\frac{G M }{\sin^2\frac{\theta}{2}}\right)^2\cos^2\frac{\theta}{2}\, \sin\theta
\label{semic1}
\end{eqnarray}
that in the small $\alpha$ (i.e. large $b$) limit takes the asymptotic form  
\begin{equation}
b = G M\left(\frac{4}{\alpha} +\frac{\alpha}{3}(1+\ln\,8-3\ln\alpha)\right) + {\cal O}(\alpha^2),
\label{blocal}
\end{equation}
which allows us to identify the deflection angle as $\alpha\sim 4 {G M}/{b}$ for large $b$ (or $b_h$), in agreement with Einstein's prediction for the angular deflection. If we consider the complete epxression of the related cross section \eqref{sigmaOL}, the expansion of the quantum corrections are organised in terms of the angle of deflection $\alpha$ and of coefficients derived from the quantum corrections. In the neutrino case they provide the relation 
\begin{equation}
\label{btheta}
b_h(E, \alpha) = \frac{2}{\alpha}+c(E)\,\alpha+d(E)\,\alpha\,\ln(\alpha)+f(E)\,\alpha^3+g(E)\,\alpha^3\ln \alpha+h(E)\,\alpha^3\ln^2 \alpha + \mathcal{O}(\alpha^5)
\end{equation}
 that we can invert in order to get the angle of deflection in terms of the energy of the incoming neutrino and the impact parameter $\alpha(E, b_h)$
\begin{equation}
\alpha(E, b_h)=\frac{2}{b_h}-\frac{1}{b_h^3}\Big[\big(2+4\,C_1(E)\big)\log b_h + \mathcal{A}(E)\Big] + \mathcal{O}(1/b_h^5).
\end{equation}
whose explicit expression can be found in \cite{coriano2015neutrino}.
\subsection{The photon case}
In the photon case, the relevant interaction is the one-loop $TVV$, with one stress energy tensor coupled to the external gravitational fluctuations $h^{\mu\nu}$, and two photon currents.   
$V^{\mu\nu\alpha\beta}(p_1,p_2)$ is the Born level interaction of the graviton vertex with the two photons and 
 $\Gamma_{(1)}^{\mu\nu\alpha\beta}$ indicate the renormalized quantum corrections at one-loop computed in the complete electroweak theory. The matrix element for the scattering of a photon in an external gravitational field is given by
\begin{eqnarray}
i\, \mathcal S^{(1)}_{if} 
&=&
2\,\pi\, \delta(q_0)\, \left( \frac{\kappa\,M}{2\,|\vec{q}|^2} \right)\, \overline{S}_{\mu\nu}\, 
\left(  V^{\mu\nu\alpha\beta}(p_1,p_2) + \Gamma_{(1)}^{\mu\nu\alpha\beta}(p_1,p_2)\right) \,
A^i_\alpha(p_1)\, A^f_\beta(p_2) \nonumber \\
\label{MatEl}
\end{eqnarray}
where $A^{i,f}$ are normalized plane waves of the incoming state and final state photon, while $\bar{S}_{\mu\nu}\equiv \eta_{\mu\nu}-2 \delta^0_{\mu}\delta^0_{\nu}$. At Born level the photon cross section is then given by 
\begin{equation}
\label{eq:crsechVV}
\left.\frac{d \sigma}{d \Omega}\right|^{(0)}_\gamma= (G M)^2\cot^4 (\theta/2) \,.
\end{equation}

At this level, from Eq.\eqref{semic} we obtain the differential equation
\begin{equation}
\frac{d b_0^2}{d\theta}= - 2\,  \left(G\,M\right)^2\, \cot^4\left(\frac{\theta}{2}\right)\, \sin\theta\, ,
\label{SquaredImpact}
\end{equation}
with $b_0$ denoting the value of $b$ computed at this order. The equation is separable and determines $b_0$ as a function of $\alpha$, modulo an integration constant.  If we set this constant to zero we obtain the solution 
\begin{equation}
b_0^2(\alpha)= 4\, G^2 M^2\, \left( \csc^2\left(\frac{\alpha}{2}\right) + 4\, \log\sin\left(\frac{\alpha}{2}\right)
  \label{bvt}                                                   - \sin^2\frac{\alpha}{2}       \right)\, .
\end{equation}
In the small $\alpha$ limit (i.e. for a large impact parameter) the solution above becomes
\begin{equation}
b_0 \sim G M\, \left(\frac{4}{\alpha} +  \frac{\alpha}{6} \left( 1 + 12\,\log \frac{\alpha}{2} \right) \right ),\,
\label{blocal}
\end{equation}
which allows us to identify the deflection angle $\alpha$ as $\alpha \sim {4 \,G\, M}/{b_0}$,
in agreement with the classical GR result. The remaining corrections arise from the Born-level result but are not dependent on energy.  A discussion of the energy dependence of the semiclassical deflection, solving for the impact parameter as a function  of $\alpha$ can be found in \cite{coriano2015neutrino}. Here we report the simpler expression obtained analyzing the post newtonial corrections to the Born level result, which become energy dependent  
\begin{align}
\left.\alpha\right|_{\gamma, \text{1PN}}^{(0)}=&\frac{2}{b_h}-\frac{1}{b_h^2}\frac{\pi}{2}E\,(GM)-\frac{1}{b_h^3}\Big(\ln b_h\big(4-\frac{1}{16}\,\pi^2E^2(GM)^2\big)-\frac{1}{64}\,\pi^2E^2(GM)^2\nonumber\\
&-\frac{4}{3}\,\pi\,E\,(GM)-\frac{1}{3}\Big) + \mathcal O(b_h^4).
\end{align}
These corrected expressions can be inserted into the lens equation (see \figref{geo1}). The equation takes the scalar form 
\begin{equation}
\label{thin1}
\beta=\theta_I- \alpha(E) \frac{D_{LS}}{D_{OS}}, 
\end{equation}
which can be extended to the case of stronger lensing by the inclusion of the contributions of the $1/b^n$ corrections 
contained in $\alpha(E)$. $D_{LS}$ and $D_{OS}$ denote the distance between the lens and the source, and the source and the observer, respectively.
\section{Conclusions}
 In conclusion, we note that this approach can be extended to other lensing observables, such as lensing magnification and others, where the deflection angle plays a direct or indirect role, including the analysis of Shapiro time delays.\\

\centerline{\bf Acknowledgements}
This work is partially funded by the European Union, Next Generation EU, PNRR project "National Centre for HPC, Big Data and Quantum Computing", project code CN00000013; by INFN, inziativa specifica {\em QG-sky} and by the grant PRIN 2022BP52A MUR "The Holographic Universe for all Lambdas" Lecce-Naples.

\vfill

\begin{thebibliography}{00}

\bibitem{Coriano:2014gia}
C.~Corian\`o, L.~Delle Rose, M.~M. Maglio, and M.~Serino.
\newblock Electroweak corrections to photon scattering, polarization and
  lensing in a gravitational background and the near horizon limit.
\newblock {\em JHEP}, 01:091, 2015.
\newblock \href{https://doi.org/10.1007/JHEP01(2015)091}{doi:10.1007/JHEP01(2015)091}.
\newblock \href{https://arxiv.org/abs/1411.2804}{arXiv:1411.2804 [hep-ph]}.

\bibitem{coriano2015neutrino}
C.~Corian\`o, A.~Costantini, M.~Dell'Atti, and L.~Delle Rose.
\newblock Neutrino and photon lensing by black holes: Radiative lens equations and
  post-newtonian contributions.
\newblock {\em JHEP}, 07:160, 2015.
\newblock \href{https://doi.org/10.1007/JHEP07(2015)160}{doi:10.1007/JHEP07(2015)160}.
\newblock \href{https://arxiv.org/abs/1504.01322}{arXiv:1504.01322 [hep-ph]}.

\bibitem{Coriano:2013iba}
C.~Corian\`o, L.~Delle Rose, E.~Gabrielli, and L.~Trentadue,
\href{https://doi.org/10.1007/JHEP03(2014)136}{``Fermion Scattering in a Gravitational Background: Electroweak Corrections and Flavour Transitions,''}
\textit{JHEP} \textbf{03}, 136 (2014)
\href{https://arxiv.org/abs/1312.7657}{[arXiv:1312.7657 [hep-ph]]}.


\bibitem{Delbourgo:1973xe}
R.~Delbourgo and P.~Phocas-Cosmetatos,
\newblock Phys.Lett. {\bf B41}, 533 (1972).

\bibitem{Berends:1974gk}
F.~A. Berends and R.~Gastmans,
\newblock Nucl. Phys. {\bf B88}, 99 (1975).

\bibitem{Schneider1992}
P.~Schneider, J.~Ehlers, and E.~E. Falco, 
\newblock\emph{Gravitational Lenses}, Astronomy and Astrophysics Library, Springer-Verlag, Berlin, Heidelberg (1992).

\end{thebibliography}
\end{document}